\documentclass[12pt]{article}
\usepackage{epsfig}
\usepackage{hyperref}%

\date{}
\begin{document}

\newcommand{\beq}{\begin{equation}}
\newcommand{\eeq}{\end{equation}}
\newcommand{\nn}{\nonumber}
\newcommand{\bea}{\begin{eqnarray}}
\newcommand{\eea}{\end{eqnarray}}

\title{On the Stability of Randall-Sundrum Braneworlds with Conformal
  Bulk Fields}

\author{Rui Neves\\
{\small \em Departamento de F\'{\i}sica, Faculdade de Ci\^encias e Tecnologia,
Universidade do Algarve}\\
{\small \em Campus de Gambelas, 8000-117 Faro, Portugal}\\
{\small E-mail: \tt rneves@ualg.pt}
}

\maketitle

\begin{abstract}

In the Randall-Sundrum scenario we consider exact 
5-dimensional solutions with localized gravity which are associated
with a well defined class of conformal bulk fields. We analyze their
behaviour under radion field perturbations. We show that if the 
Randall-Sundrum exponential warp is the localizing metric 
function and the equation of state of the conformal fields
is not changed by the radion perturbation then the 5-dimensional solutions
are unstable. We present new stable solutions which describe on the
brane the dynamics of inhomogeneous dust, generalized dark radiation
and homogeneous polytropic matter.   

\end{abstract}
\section{Introduction}

In the Randall-Sundrum (RS) scenario \cite{RS1,RS2} the
observable Universe is a 3-brane world of a $Z_2$ symmetric 
5-dimensional anti-de Sitter (AdS) space. In the RS1 model
\cite{RS1} the AdS orbifold has a compactified fifth dimension and two brane
boundaries. The gravitational field is bound to the hidden positive
tension brane and decays towards the visible negative tension brane. 
In this model the 
hierarchy problem is reformulated as an exponential hierarchy between the weak
and Planck scales \cite{RS1}. In the RS2 model \cite{RS2} the AdS
orbifold is non-compact with an
infinite fifth dimension and a single positive tension brane. 
Gravity is localized on the positive tension brane now
interpreted as the visible brane. 

At low energies the theory of gravity on the observable brane is 
4-dimensional general relativity and the cosmology may be 
Friedmann-Robertson-Walker \cite{RS1}-\cite{TM}. 
In the RS1 model this is only possible if  
the radion mode is stabilized and this as been achieved using a scalar
field in the bulk \cite{GW,WFGK,CGRT,TM}. The gravitational 
collapse of matter was also analyzed in the RS scenario \cite{CHR}-\cite{RC2}. 
Using an extended black string
solution (first discussed in a different context by Myers and Perry
\cite{MP}) Chamblin, Hawking and Reall showed that it was possible to
induce on the brane the Schwarzschild
black hole metric \cite{CHR}. However, this solution is divergent at 
the AdS horizon and at
the black string singularity. Consequently, 
it is expected to be unstable \cite{CHR,GL}. A black cylinder localized near the brane which is free from naked 
singularities was conjectured to be its stable decay product. This exact 5-dimensional solution has not yet been
found. The only known static black holes localized
on a brane remain to be those found for a 2-brane in a 4-dimensional AdS space
\cite{EHM}. The problem lies in the simultaneous non-singular localization of
gravity and matter in the vicinity of the
brane \cite{CHR}, \cite{KOP}-\cite{RC2}. This has lead to another conjecture
stating that $D+1$-dimensional black hole solutions localized on
a $D-1$-brane should correspond to quantum corrected $D$-dimensional
black holes on the brane \cite{EFK}. This is an extra motivation to
look for 5-dimensional collapse solutions localized on a brane. In
addition, the effective covariant
Gauss-Codazzi approach \cite{BC,SMS} has permitted the discovery of
many braneworld solutions which have not yet been associated with
exact 5-dimensional spacetimes 
\cite{COSp}-\cite{RC1}.

In this paper we continue the research
on the dynamics of a spherically symme\-tric RS 3-brane when the bulk
is filled with conformal matter fields \cite{RC2,RC3} (see also
\cite{EONO}). In our previous work
\cite{RC2,RC3} we have found a new class of exact
5-dimensional dynamical solutions for which gravity is localized near
the brane by the exponential RS warp. These solutions were shown to be
associated with conformal bulk fields characterized by a stress-energy
tensor $\tilde{T}_\mu^\nu$ of weight -4 and by the equation of
state ${\tilde{T}_a^a}=2{\tilde{T}_5^5}$. They were also shown
to describe on the brane the dynamics of
inhomogeneous dust, generalized dark radiation and homogeneous
polytropic matter. However, the density and pressures
of the conformal bulk fluid increase with the coordinate of the fifth
dimension. In the RS2 model this generates a divergence at the AdS
horizon as in the Schwarzschild black string solution. In
the RS1 model this is not a difficulty because the space is cut before the AdS horizon is reached. However,
the solutions must be stable under radion field perturbations. In this
paper we analyze this problem using a saddle point expansion procedure
based on the action \cite{HKP,CGHW}. In section 2 we present a brief
review of the conformal analysis performed on the 5-dimensional Einstein
equations which leads to the set of braneworld backgrounds 
supported by the class of conformal bulk matter fields with weight
-4. In section 3 we consider a
dimensional reduction on the RS action and determine the radion
effective potential. In section 4 we identify our 
braneworld solutions as its extrema and show 
that the exponential RS warp and a radion independent equation of
state for the conformal bulk fields lead to unstable solutions. In
section 5 we present our conclusions and a new class of stable
braneworld solutions which describe on the brane the dynamics of
inhomogeneous dust, generalized dark radiation and homogeneous
polytropic matter. 

\section{Einstein Equations and Conformal Bulk Fields}  

Let $(t,r,\theta,\phi,z)$ be a set of comoving coordinates in the RS 
orbifold. The most general metric consistent with the 
$Z_2$ symmetry in $z$ and with 4-dimensional spherical symmetry on the
brane may be written as  

\beq
d{\tilde{s}^2}={\Omega^2}\left(-{e^{2A}}d{t^2}
+{e^{2B}}d{r^2}+{R^2}d{\Omega_2^2}+d{z^2}\right),
\eeq
where the metric functions $\Omega=\Omega(t,r,z)$, $A=A(t,r,z)$, 
$B=B(t,r,z)$ and $R=R(t,r,z)$ are $Z_2$ symmetric. $\Omega$ is the
warp factor and $R$ the physical radius of the 2-spheres.

When the 5-dimensional space is filled with bulk matter fields
characterized by a lagrangian $\tilde{L}_B$ the dynamical
RS action is given by

\beq
{\tilde{S}}=\int{d^4}xdz\sqrt{-\tilde{g}}
\left\{{\tilde{R}\over{2{\kappa_5^2}}}-
{\Lambda_B}-{1\over{\sqrt{\tilde{g}_{55}}}}\left[\lambda\delta\left(z-{z_0}\right)+\lambda'\delta\left(z-{{z'}_0}\right)\right]+
{\tilde{L}_B}\right\}.\label{5Dact}
\eeq
The Planck brane with tension $\lambda$ is assumed to be located at $z={z_0}$ and the visible
brane with tension $\lambda'$ at ${z'}_0$. $\Lambda_B$ is the negative bulk cosmological constant. A Noether variation on the action (\ref{5Dact}) gives the 
Einstein field equations

\beq
{\tilde{G}_\mu^\nu}=-{\kappa_5^2}\left\{{\Lambda_B}{\delta_\mu^\nu}+
{1\over{\sqrt{\tilde{g}_{55}}}}\left[\lambda\delta
\left(z-{z_0}\right)+\lambda'\delta
\left(z-{{z'}_0}\right)\right]\left(
{\delta_\mu^\nu}-{\delta_5^\nu}{\delta_\mu^5}\right)-
{\tilde{T}_\mu^\nu}\right\},
\label{5DEeq}
\eeq
where the stress-energy tensor associated with the bulk fields is defined by

\beq
{\tilde{T}_\mu^\nu}={\tilde{L}_B}{\delta_\mu^\nu}-
2{{\delta{\tilde{L}_B}}\over{\delta{\tilde{g}^{\mu\alpha}}}}
{\tilde{g}^{\alpha\nu}}
\eeq
and is conserved in the bulk,
${\tilde{\nabla}_\mu}{\tilde{T}_\nu^\mu}=0.$ 

To find exact solutions of the 5-dimensional Einstein equations
(\ref{5DEeq}) we need simplifying assumptions. 
Let us first consider that under the 
conformal transformation ${\tilde{g}_{\mu\nu}}={\Omega^2}{g_{\mu\nu}}$
the bulk stress-energy tensor has conformal weight $s$,
${\tilde{T}_\mu^\nu}={\Omega^{s+2}}{T_\mu^\nu}$. Then
Eq. (\ref{5DEeq}) may be re-written as 

\[
{G_\mu^\nu}=-6{\Omega^{-2}}\left({\nabla_\mu}\Omega\right){g^{\nu\rho}}
{\nabla_\rho}\Omega+
3{\Omega^{-1}}{g^{\nu\rho}}{\nabla_\rho}{\nabla_\mu}\Omega
-3{\Omega^{-1}}{\delta_\mu^\nu}{g^{\rho\sigma}}{\nabla_\rho}{\nabla_\sigma}
\Omega
\]
\beq
-{\kappa_5^2}
{\Omega^2}\left\{{\Lambda_B}{\delta_\mu^\nu}+{\Omega^{-1}}\left[\lambda
\delta(z-{z_0})+\lambda'\delta(z-{{z'}_0})\right]\left(
{\delta_\mu^\nu}-{\delta_5^\nu}{\delta_\mu^5}\right)
-{\Omega^{s+2}}{T_\mu^\nu}\right\}.\label{t5DEeq}
\eeq
Similarly under the conformal transformation the conservation equation becomes

\beq
{\nabla_\mu}{T_\nu^\mu}+{\Omega^{-1}}\left[(s+7){T_\nu^\mu}{\partial_\mu}
\Omega-{T_\mu^\mu}{\partial_\nu}\Omega\right]=0\label{t5Dceq}.
\eeq
If in addition it is assumed that
${{\tilde{T}}_\mu^\nu}={\Omega^{-2}}{T_\mu^\nu}$ then
equation (\ref{t5DEeq}) may be separated in the following way

\beq
{G_\mu^\nu}={\kappa_5^2}{T_\mu^\nu},\label{r5DEeq}
\eeq
\[
6{\Omega^{-2}}{\nabla_\mu}\Omega{\nabla_\rho}
\Omega{g^{\rho\nu}}-
3{\Omega^{-1}}{\nabla_\mu}{\nabla_\rho}\Omega{g^{\rho\nu}}+3{\Omega^{-1}}
{\nabla_\rho}{\nabla_\sigma}\Omega{g^{\rho\sigma}}{\delta_\mu^\nu}=
\]
\beq
-{\kappa_5^2}
{\Omega^2}\left\{{\Lambda_B}{\delta_\mu^\nu}+{\Omega^{-1}}\left[
\lambda\delta(z-{z_0})+\lambda'\delta(z-{{z'}_0})\right]\left(
{\delta_\mu^\nu}-{\delta_5^\nu}{\delta_\mu^5}\right)\right\}.\label{5DEeqwf}
\eeq
Because of the Bianchi identity we must also have

\beq
{\nabla_\mu}{T_\nu^\mu}=0,\label{r5Dceq}
\eeq
\beq
3{T_\nu^\mu}{\partial_\mu}\Omega-T {\partial_\nu}\Omega=0.
\label{5Dceqwf}
\eeq
Equations (\ref{r5DEeq}) and (\ref{r5Dceq}) are 5-dimensional
Einstein equations with matter fields present in the bulk but without
a brane or bulk cosmological constant. They do not depend on the conformal
warp factor which is dynamically defined by equations (\ref{5DEeqwf}) and 
(\ref{5Dceqwf}). Consequently, only the warp reflects the
existence of the brane or of the bulk cosmological constant. Note that
this is only possible for the special class of conformal bulk fields
which have a stress-energy tensor with weight -4.

Because the equations still depend non-linearly on the metric 
functions $A$, $B$ and $R$ let us further assume that $A=A(t,r)$, $B=B(t,r)$,
$R=R(t,r)$ and $\Omega=\Omega(z)$. Then we obtain 

\beq
{G_a^b}={\kappa_5^2}{T_a^b},\quad{\nabla_a}{T_b^a}=0,\label{4DEeq}
\eeq
\beq
{G_5^5}={\kappa_5^2}{T_5^5},\label{5DEeqz}
\eeq
\beq
6{\Omega^{-2}}{{({\partial_z}\Omega)}^2} =
-{\kappa_5^2}{\Omega^2}{\Lambda_B},
\eeq
\beq
3{\Omega^{-1}}{\partial_z^2}\Omega = -{\kappa_5^2}{\Omega^2}
\left\{{\Lambda_B}+{\Omega^{-1}}\left[\lambda\delta(z-{z_0})+\lambda'
\delta(z-{{z'}_0})\right]\right\}\label{rswf}
\eeq
and (see also \cite{KKOP} and \cite{IR})
\beq
2{T_5^5}={T_c^c},\label{eqst1}
\eeq
where the latin indices represent the coordinates $t,r,\theta$ and
$\phi$. Our braneworld geometries \cite{RC2,RC3} are solutions of equations
      (\ref{4DEeq})-(\ref{eqst1}) when the stress-tensor is diagonal,
\beq
{T_\mu^\nu}=diag\left(-\rho,{p_r},{p_T},{p_T},{p_5}\right),\label{bmten}
\eeq
where $\rho$, $p_r$, $p_T$ and $p_5$ denote the bulk matter density and
pressures, and $\Omega$ is the
      exponential RS warp, 

\beq
{\Omega_{\mbox{\tiny
      RS}}}(z)={l\over{|z-{z_0}|+{z_0}}},
\eeq 
where ${z_0}=l$ and $l$ is the AdS radius given by $l=1/\sqrt{-{\Lambda_B}{\kappa_5^2}/6}$ with 
${\kappa_5^2}=8\pi/{M_5^3}$ defined by the fundamental 5-dimensional 
Planck mass $M_5$. Then the Planck brane is located at ${z_0}=l$ and the 
observable brane is located at ${{z'}_0}=l{e^{\pi{r_c}/l}}$ where
      $r_c$ is the RS compactification scale \cite{RS1}. The former
      has a positive tension $\lambda$ and the latter a negative
      tension $\lambda'=-\lambda$ where $\lambda=-{\Lambda_B}l$. 
They are twin Universes with identical collapse or cosmological
      dynamics.    

\section{The Radion Potential}

To analyze the behaviour of these solutions under radion field
perturbations we apply a saddle point expansion procedure
based on the action \cite{HKP,CGHW}. As a starting point this requires 
the determination of the radion effective potential. For the calculation it is
convinient to work with the coordinate $y$ related to $z$ by
$z=l{e^{y/l}}$ for $y>0$. Let us write the most general metric
consistent with the 
$Z_2$ symmetry in $y$ and with 4-dimensional spherical symmetry on the
brane in the form

\beq
d{\tilde{s}^2}={a^2}d{s_4^2}+{b^2}d{y^2},\quad d{s_4^2}=-d{t^2}
+{e^{2B}}d{r^2}+{R^2}d{\Omega_2^2},\label{5Dmt}
\eeq
where the metric functions $a=a(t,r,y)$, $B=B(t,r,y)$,
$R=R(t,r,y)$ and $b=b(t,r,y)$ are $Z_2$ symmetric. Now $a$ is the warp
factor, $R$ is still the physical radius of the 2-spheres and $b$ is related
to the radion field. 

The 5-dimensional dynamical RS action is now given by

\beq
\tilde{S}=\int{d^4}xdy\sqrt{-\tilde{g}}
\left\{{\tilde{R}\over{2{\kappa_5^2}}}-
{\Lambda_B}-{1\over{\sqrt{\tilde{g}_{55}}}}\left[\lambda\delta\left(y\right)+\lambda'\delta\left(y-\pi
  {r_c}\right)\right]+
{\tilde{L}_B}\right\}.\label{5Dact1}
\eeq
In the new coordinates the Planck brane is located at $y=0$ and the visible
brane at $\pi {r_c}$. Our braneworld
backgrounds correspond to the metric functions $b=1$,
$B=B(t,r)$, $R=R(t,r)$ and $a={\Omega_{\mbox{\tiny
      RS}}}(y)$ where 

\beq
{\Omega_{\mbox{\tiny
      RS}}}(y)={e^{-|y|/l}}. 
\eeq

To calculate the radion potential we consider the dimensional
reduction of the action (\ref{5Dact1}). Using the metric (\ref{5Dmt}) we obtain
$\sqrt{-\tilde{g}}={a^4}b\sqrt{-{g_4}}$ and  

\[
\tilde{R}={1\over{a^2}}\left({R_4}-{6\over{a}}{g_4^{cd}}{\nabla_c}{\nabla_d}a
-{2\over{b}}{g_4^{cd}}{\nabla_c}{\nabla_d}b-{4\over{ab}}{g_4^{cd}}{\nabla_c}a{\nabla_d}b\right)
\]
\beq
-{{4}\over{b^2}}\left[3{{\left({{{\partial_y}a}\over{a}}\right)}^2}+2{{{\partial_y^2}a}\over{a}}\right],
\eeq 
where ${g_4^{cd}}$ is the inverse metric associated with the
4-dimensional line element $d{s_4^2}$, $R_4$ is the 4-dimensional Ricci
scalar and the covariant derivatives are 4-dimensional. Consider the particular metric setting defined by 
$a=\Omega{e^{-\beta}}$ and $b={e^\beta}$ where $\Omega=\Omega(y)$ and 
$\beta=\beta(t,r)$. Then in the Einstein frame the dimensional
reduction leads to 

\beq
\tilde{S}=\int{d^4}x\sqrt{-{g_4}}\left({{R_4}\over{2{\kappa_4^2}}}-{1\over{2}}{\nabla_c}\gamma{\nabla_d}\gamma{g_4^{cd}}-\tilde{V}\right),\label{DR5Dact}
\eeq
where $\gamma=\beta/({\kappa_4}\sqrt{2/3})$ is the canonically
normalized radion field. The function $\tilde{V}=\tilde{V}(\gamma)$ is the radion 
potential and it may be written in the form

\[
\tilde{V}={2\over{\kappa_5^2}}{\chi^3}\left[3
\int dy{\Omega^2}
{{({\partial_y}\Omega)}^2}+2\int dy{\Omega^3}
{\partial_y^2}\Omega\right]+\chi\int dy{\Omega^4}\left({\Lambda_B}-{\tilde{L}_B}\right)
\]
\beq
+{\chi^2}\int dy{\Omega^4}\left[\lambda\delta\left(y\right)+\lambda'\delta\left(y-\pi
  {r_c}\right)\right],\label{rp}
\eeq
where the field $\chi$ is defined as
$\chi={e^{-\sqrt{(2/3)}\;{\kappa_4}\gamma}}$. Note that the integration 
in the fifth dimension is performed in the interval
$\left[-\pi{r_c},\pi{r_c}\right]$ and that we have chosen 

\beq
\int dy{\Omega^2}={{\kappa_5^2}\over{\kappa_4^2}}.\label{nc}
\eeq

\section{The Radion Field Instability}

To analyze the stability of our braneworld solutions we consider a
saddle point expansion of the radion field potential $V$
\cite{HKP,CGHW}. This procedure requires the
determination of its first and second variations. To do so we consider 
the integral of the radion potential written in the form

\[
\tilde{\mathcal{V}}=\int{d^4}x\sqrt{-{g_4}}\tilde{V}(\gamma)=\int{d^5}x\sqrt{-\tilde{g}}\left\{{2\over{\kappa_5^2}}{\chi^2}\left[3{{\left({{{\partial_y}\Omega}\over{\Omega}}\right)}^2}+2{{{\partial_y^2}\Omega}\over{\Omega}}\right]\right\}
\]
\beq
+\int{d^5}x\sqrt{-\tilde{g}}\left\{{\Lambda_B}-{\tilde{L}_B}+\chi\left[\lambda\delta\left(y\right)+\lambda'\delta\left(y-\pi
  {r_c}\right)\right]\right\}.
\eeq     
Taking into account that an integration by parts and the $Z_2$
symmetry leads to 

\beq
\int dy{\Omega^2}{{({\partial_y}\Omega)}^2}=-{1\over{3}}\int dy{\Omega^3}{\partial_y^2}\Omega
\eeq
we find that the first variation of the integral of the radion
potential is given by

\[
{{\delta \tilde{\mathcal{V}}}\over{\delta\gamma}}=-\sqrt{8\over{3}}\;{\kappa_4}\int{d^5}x\sqrt{-\tilde{g}}\chi\left[\lambda\delta\left(y\right)+\lambda'\delta\left(y-\pi
  {r_c}\right)\right]-\sqrt{2\over{3}}\;{\kappa_4}\int{d^5}x\sqrt{-\tilde{g}}{\Lambda_B}
\]
\beq
+{{\kappa_4}\over{\sqrt{6}}}\int{d^5}x\sqrt{-\tilde{g}}\left({\tilde{T}_a^a}-2{\tilde{T}_5^5}-{{12}\over{\kappa_5^2}}{\chi^2}{{{\partial_y^2}\Omega}\over{\Omega}}\right).\label{fviV}
\eeq
For our solutions $\Omega$ is taken to be the exponential RS warp
factor $\Omega_{\mbox{\tiny
    RS}}$ and satisfies the following warp equation in the $y$ coordinate 

\beq
-{{6}\over{\kappa_5^2}}{{{\partial_y^2}\Omega}\over{\Omega}}={\Lambda_B}+2\left[\lambda\delta\left(y\right)+\lambda'\delta\left(y-\pi
  {r_c}\right)\right]\label{we1}.
\eeq
On the
other hand the bulk matter fields which must have a stress-energy tensor of
conformal weight -4 must also obey the equation of
state (\ref{eqst1}). 
Integrating in the fifth dimension with $\Omega={\Omega_{\mbox{\tiny
    RS}}}$ we find

\beq
{{\delta \tilde{V}}\over{\delta\gamma}}={{\kappa_4}\over{\sqrt{6}}}{\Lambda_B}l\left(1-{e^{-4\pi{r_c}/l}}\right)\chi\left(4\chi-1-3{\chi^2}\right).
\eeq
The critical extrema of the radion potential are the non-zero finite
roots of the polynomial equation $\chi(3{\chi^2}-4\chi+1)=0$. They are 
${\chi_1}=1$ and ${\chi_2}=1/3$. Our braneworld solutions correspond
to the first root ${\chi_1}=1$. The same happens if the bulk matter
is absent as in the RS vaccum solutions. The other extremum
is not a solution of the Einstein equations for the RS1
model. Indeed, a rescaling of the coordinates shows that the point
${\chi_2}=1/3$ corresponds to an exponential warp
$\Omega={e^{-|y|/(3l)}}$ which is different 
from $\Omega_{\mbox{\tiny
    RS}}$ and does not satisfy the RS warp
equations (\ref{we1}) and  

\beq     
6{{\left({{{\partial_y}\Omega}\over{\Omega}}\right)}^2}=-{\kappa_5^2}{\Lambda_B}.
\eeq
The stability of the extrema depends on the sign of the second
variation of the radion potential. This variation defines the
radion mass. Consequently, stable background solutions must be associated 
with a positive sign. Consider the first variation of the
4-dimensional integral of $V$ given in Eq. (\ref{fviV}). If the equation of
state (\ref{eqst1}) of the conformal bulk fields is independent of the
radion perturbation and $\Omega={\Omega_{\mbox{\tiny
    RS}}}$ we find 

\beq
{{{\delta^2}V}\over{\delta{\gamma^2}}}={{\kappa_4^2}\over{3}}{\Lambda_B}l\left(1-{e^{-4\pi{r_c}/l}}\right)\chi\left(9{\chi^2}-8\chi+1\right).
\eeq
For $\chi={\chi_1}=1$ we conclude that the second variation of
$V$ is negative. This implies that the radion is a negative
mass tachyon and that our braneworld solutions are unstable. 

\section{Conclusions}

In this paper we have considered the set of exact
5-dimensional dynamical solutions with gravity localized near
the brane which are associated with conformal bulk fields of weight -4
and which describe the dynamics of
inhomogeneous dust, generalized dark radiation and homogeneous
polytropic matter on the brane. We have studied their behviour under
radion field perturbations. We have shown that these solutions are
extrema of the radion potential. We have also shown
that if the metric function responsible for the localization of gravity
is the exponential RS warp and the equation of state
characterizing the conformal bulk fluid is independent of the radion field 
then the braneworld solutions are unstable. We have also found that
the radion potential has another extremum which is not a solution of
the complete set of Einstein equations. This point is connected with a
different warp factor and its existence suggests that stable
braneworlds associated with the same state of the conformal
bulk matter should correspond to warp functions other than the
standard exponential RS warp. This is indeed true. Stable solutions
are defined by a new set of warp functions given by

\beq
\Omega(y)={e^{-|y|/l}}\left(1+{{p_5^s}\over{4{\Lambda_B}}}{e^{2|y|/l}}\right),
\eeq
where ${p_5^s}<0$ is a negative constant fraction of the 
5-dimensional pressure of the conformal bulk fields of weight
-4. On the brane these solutions also describe the dynamics of
inhomogeneous dust, generalized dark radiation and homogeneous
polytropic matter. More details will be presented in a forthcoming 
publication \cite{RC4}.   
\vspace{0.5cm}

\centerline{\bf Acknowledgements}
\vspace{0.5cm}

We are grateful for financial support from {\it Funda\c {c}\~ao para a Ci\^encia e a Tecnologia} (FCT) and {\it Fundo Social Europeu} (FSE) under the contracts
SFRH/BPD\-/7182/2001 and POCTI/32694/FIS/2000
({\it III Quadro Comunit\'ario de Apoio}) as well as from {\it Centro
  Multidisciplinar de Astrof\'{\i}sica} (CENTRA). We would also like to thank
Cenalo Vaz for enlighting discussions and to the Physics Department 
of the University of Cincinnati for kind hospitality during a visit 
where part of this work was completed.

\end{document}